\documentclass[12pt,oneside,a4paper]{article}
\usepackage{amsfonts}  
\usepackage{amsmath}   
\usepackage[T1]{fontenc}
\usepackage[cp1250]{inputenc}
\usepackage{graphicx}
\graphicspath{ {./images/} }
\usepackage{bm}
\usepackage{tikz}
\usepackage{subcaption}
\usepackage{float}
\usepackage{amsmath}
\usepackage{url}
\usepackage{hyperref}

\usepackage[top=3.5cm,bottom=2cm,right=2cm,left=3.5cm]{geometry}  
\begin{document}


\includegraphics[scale=0.6]{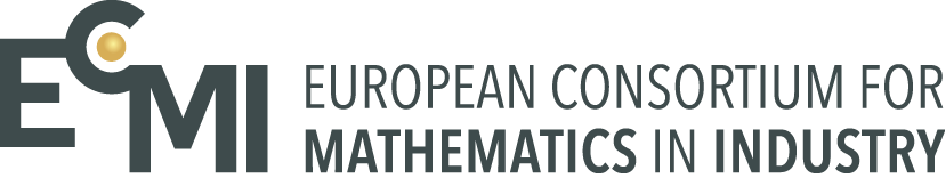}

Winter ECMI Modelling Week 2019\\
Technische Universit\"at Darmstadt

\begin{center}
\vspace{2.0cm}
{\Large{\textbf{Hybrid Modelling Approaches \\ for Forecasting Energy Spot Prices}}}

\vspace{\stretch{0.01}}
Instructor: Prof.\ Dr.\ Matthias Ehrhardt (Bergische Universit\"at Wuppertal)
\vspace{\stretch{0.01}}

Authors:

\begin{tabular}{c}
Tahir Miriyev (University of Verona) \\ 
Alessandro Contu (University of Verona) \\ 
Kevin Sch\"{a}fers (Bergische Universit\"{a}t Wuppertal) \\  
Ion Gabriel Ion (Technische Universit\"{a}t Darmstadt)
\end{tabular}

\vspace{\stretch{0.5}}

March 29, 2019
\end{center}

\vspace{\stretch{0.15}}

\newpage
\tableofcontents


\newpage

\section{Introduction}

\begin{figure}[H]
\centering \hspace{-1cm}		\includegraphics[width=0.9\textwidth]{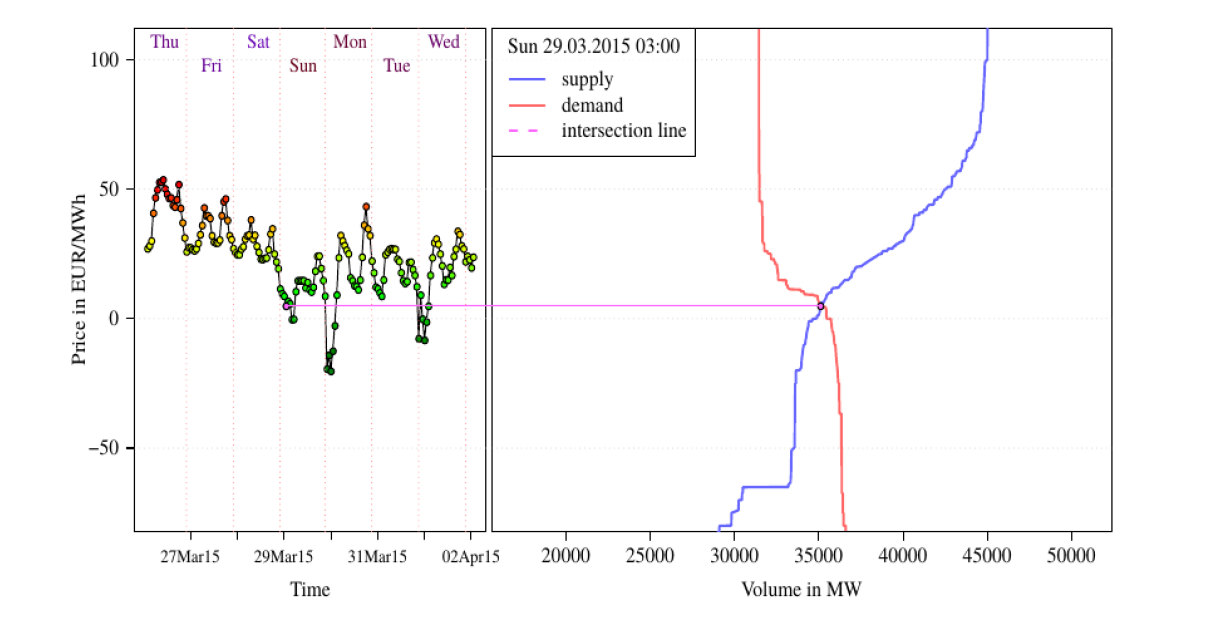}
\caption{EPEX day-ahead electricity prices and corresponding supply/sale and demand/purchase curves.}\label{fig:1}
\end{figure}

In the following sections we will discuss several methods applied for forecasting energy spot prices, in particular, electricity spot prices (ESP) for EPEX market. Electricity, by its nature, is a non-storable type of commodity. For this reason, the stability of power systems requires a balance between production and consumption, since spot prices fluctuate according to varying supply and demand curves (Fig.~\ref{fig:1}). Meanwhile, electricity demand and supply values depend on several factors, such as weather temperature, on/off-peak hours, holidays, governmental policies etc. These and many other factors affect price dynamics, exhibiting different seasonalities and price strikes. Therefore, price forecasts from a few hours to a few months ahead are of particular interest to power portfolio managers. A power market company, which is able to forecast the electricity prices with some reasonable level of accuracy, can adjust its bidding strategy and consumption/production schedule in a way to maximize the profit and reduce risks in a day-ahead trading. 

Within a week we decided to investigate the structure of the ESP, design the hierarchy of different mathematical models with increasing complexity,calibrate models with real data and finally, combine models to achieve better forecasts. Since the ultimate goal was to develop a hybrid model, we decided to study simultaneously a few models from each subgroup of the fundamental group of models. Particularly, we focused on the following models:
\begin{itemize}
\item Naive model and Fourier Series approach 
\item Mean-reversion and jump-diffusion model from the Reduced-form subgroup
\item ARMA and GARCH models from the Statistical subgroup
\item Artificial Neural Network (ANN) model from the Computational Intelligence subgroup
\end{itemize}

In practice, with only one model people always experience relatively "bad" forecasts from time to time, while combining models allows us to focus on periods where these models disagree with each other significantly and build the right logic behind forecasting. Hence, having more than one model in the complete forecasting system is always a better idea. 

The report is organized as follows:
\begin{itemize}
\item in Sections~\ref{sec:naive} and \ref{sec:fourier} we describe our approaches based on Naive model and Fourier Series Analysis
\item in Section~\ref{sec:timeseries} we explains our investigations on applying ARMA and GARCH models, along with the analysis of the effect of seasonality parameter in these models 
\item in Section~\ref{sec:mrjd} we discuss the application of MRJD model
\item in Section~\ref{sec:ann} we share our results in the implementation of Artificial Neural Networks model. 
\item in Section~\ref{sec:hybrid} we will demonstrate the forecasts under the final hybrid model
\item Section~\ref{sec:conclusions} we will give a summary of  our investigations, observations, conclusions and suggestions.
\end{itemize}

\subsection{Background and Remarks}
ESP Forecasting (ESPF) can be done either in terms of point forecasting or probabilistic forecasting. 
Probabilistic forecasts simply summarize probabilities of future events. In contrast to point-forecasting, instead of predicting some exact results, probabilistic forecast assigns a probability to each possible outcome and the complete probability distribution represents the result. 
Since by its nature the forecasting is a stochastic problem where forecasts contain some randomness, we though that it will make more sense to focus on  probabilistic methods of forecasting. However later on, as we delved deeper into the theory and literature, we realized that on practice, a lot of decision making processes simply can't take probabilistic inputs and consequently, produce probabilistic outputs. 
Also, we noticed that there is much more literature on point-forecasting compared to probabilistic one. And lastly, after analyzing the performance of teams who were working on the similar project throughout four months within the frame of Global Energy Forecasting Competition 2014 \cite{3}, we realized that due to time constraints it will be extremely hard to conduct enough research on probabilistic forecasting.Therefore we decided to focus on predicting the exact electricity prices.

We initially planned to forecast ESP for all three known forecasting horizons: short, medium and long-term horizons, which stand for a few hours, days and months-ahead forecasts, respectively. We have to point out that majority of ESPF studies in literature have focused on a short-term horizon and very little work is done on medium-term price forecasting. 

Throughout the process we understood that one of the crucial aspects of ESPF is the right choice of variables after the data mining. The current spot prices always depend on many drivers such as system loads, weather variables (temp., humid., wind speed etc), reverse margin (available generation $\pm$ predicted demand), information about scheduled maintenance, forced outages and so on . In our case, we were given  the electricity spot prices for 2008-2015 years only. Unfortunately it turned out to be difficult to find the data related to above-mentioned factors, as usually power companies or private organizations do not make this kind of data publicly available, but rather sell it. So we had to use only the data of past prices and thus, didn't perform any data mining with several variables. We will come back to this question in the Discussions section \cite{4,5}.

Relying on the literature, we also considered another key point in electricity spot price forecasting, which is an appropriate treatment of  seasonality components in our models. In general, electricity prices exhibit 3 seasonality levels: daily, weakly and yearly. We will speak more about the affect of seasonality parameters in coming sections. 

Lastly, increasing the accuracy of the forecast is directly related to the calibration of parameters. 
Probably the hardest part of the forecasting process is to understand the nature of errors and to calibrate parameters in a right way.  

\section{Our approaches}
\subsection{Naive model}\label{sec:naive}
A suitable starting point for getting into the topic of forecasting energy spot prices is a naive model which is focusing on some noteworthy observations based on the weekly seasonality:
\begin{itemize}
\item lower price on the weekend
\item similar price on weekdays
\item little differences on Monday morning because the working week starts
\item little differences on Friday afternoon because the weekend starts
\end{itemize}
Considering this observations, we are able to forecast the next week by using the following formula (we denote by $Y_{d,s}$ the electricity price at day $d$ and time period $s\in \{1,\dots,24\}$ \cite{8}): 
\begin{equation*}
Y_{d,s} = 
\begin{cases} Y_{d-7,s} + \epsilon_{d,s} &, d \text{ is Saturday or Sunday} \\ 
 \frac{1}{2} (Y_{d-1,s} + Y_{d-7,s}) + \epsilon_{d,s} &, d \text{ is Friday}\\
  \frac{1}{2} (Y_{d-3} + Y_{d-7,s}) + \epsilon_{d,s} & ,d \text{ is Monday}\\
  Y_{d-1,s} + \epsilon_{d,s} & , otherwise   \end{cases}
\end{equation*}
By applying this model on our given data of 2011 we get the following plots:\\
\begin{figure}[H]
\centering \hspace{-1cm}
\includegraphics[width=0.6\textwidth]{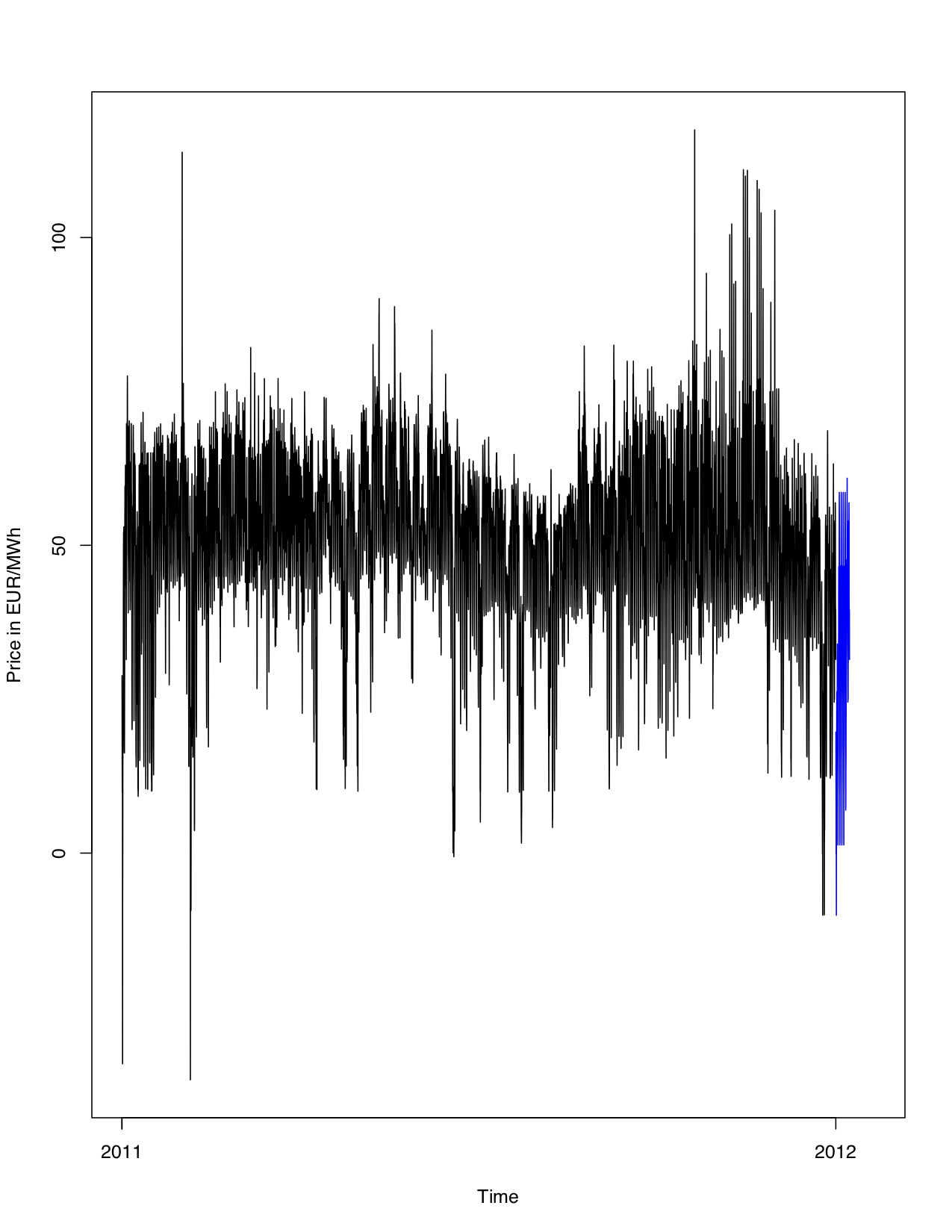}
\caption{plot of the given data (black) and the forecast (blue)}\label{fig:2}
\end{figure}

\begin{figure}[H]
\centering \hspace{-1cm}
\includegraphics[width=0.7\textwidth]{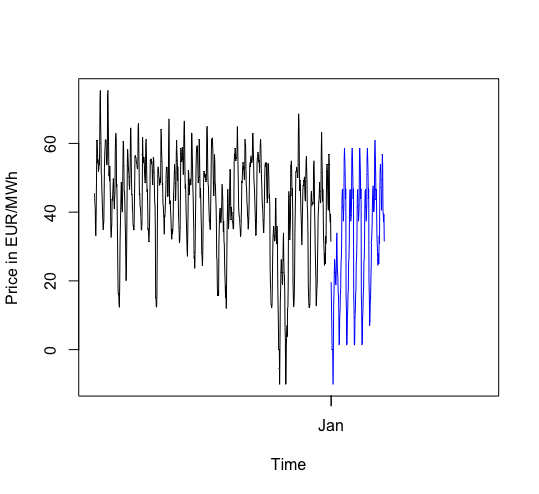}
\caption{detailed view of the forecast}\label{fig:3}
\end{figure}

To evaluate this forecast we plot the exact data of 2012 and our forecast:
\begin{figure}[H]
\centering \hspace{-1cm}
\includegraphics[width=10cm]{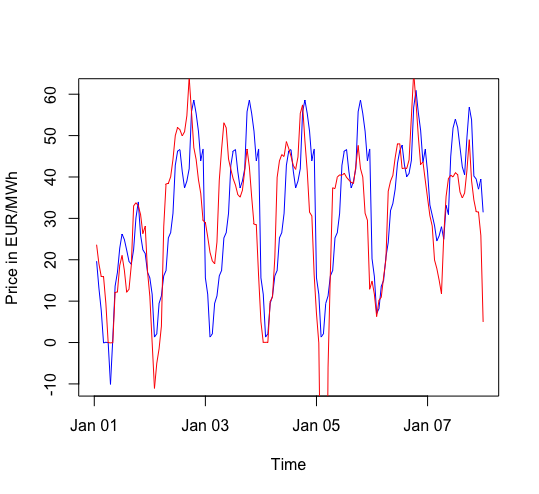}
\caption{plot of the given data (red) and the forecast (blue)}
\label{fig:4}
\end{figure}
Obviously there are bigger errors, especially at time points where the given data has outliers. 
Nevertheless this is a first good result because the plot shows the typical daily behaviour of the price. 
However this model is only useful for forecasting the next week because otherwise you would nearly copy the forecast for every week that has to be forecasted.

\noindent To create a forecast for a longer time, it is necessary to use the yearly seasonality. This is part of the next approach.

\subsection{An approach based on Fourier series}\label{sec:fourier}
The main idea of this approach is to use given data of several years to create Fourier series which show a yearly seasonality.\\
We forecast a new data point in 2015 by using
\begin{itemize}
\item Computed Fourier series
\item The given data points of the same weekdays at a similar time of the year
\item The data point at the same time of the previous week
\item A general development of recent years
\end{itemize}

\subsubsection{Computing Fourier series}
We are going to compute a Fourier series for every year (2008-2014) to forecast the stock prices in 2015. 
The results are calculated using Matlab's function \texttt{fourier8} and shown in the following figure:
\begin{figure}[H]
\centering \hspace{-1cm}
\includegraphics[width=12cm]{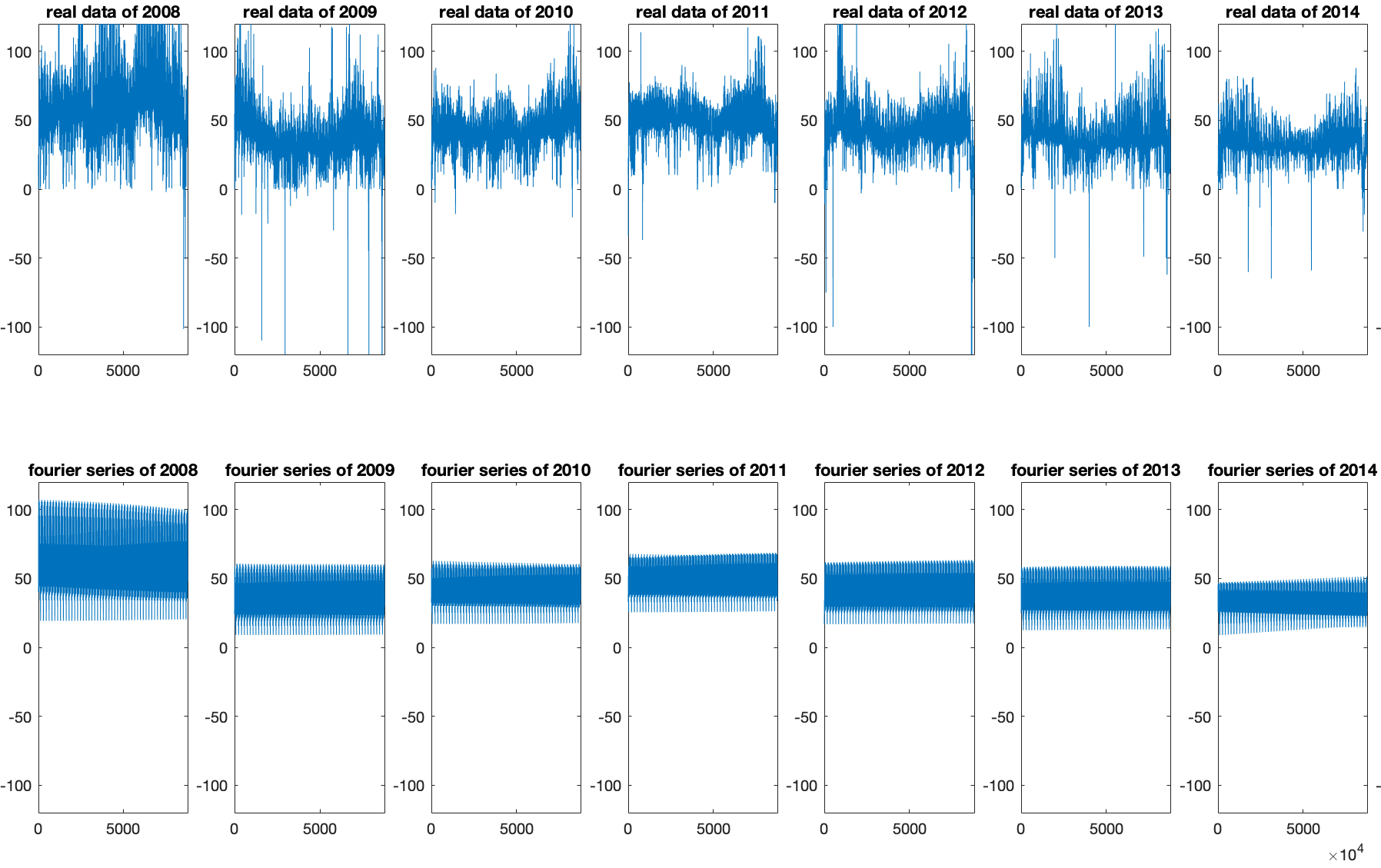}
\caption{The given data and the computed Fourier series (2008-2014)}\label{fig:5}
\end{figure}

Afterwards we calculate the average of all Fourier series where the series of four years ago (2011) and of the previous year (2014) are weighted more heavily. \\
The typical German business cycle has a duration of about four years. 
Because of this we can predict that the economic situation in 2015 will be equal to the situation in 2011. 
However there is also a kind of trend that we have to consider by weighting the data of the previous year more heavily too.
All in all we get the following formula:
\begin{align*}
z_{1} = \frac{f_{2008} + f_{2009} + f_{2010} + 10 \cdot f_{2011} + f_{2012} + f_{2013} + 5 \cdot f_{2015}}{20}
\end{align*}
with $f_{i}$ the Fourier series of the different years.

\subsubsection{Important data points}
On the one hand we want to include the given data points of the same weekdays at a similar time of the year. For calculating the next value $z_{2}$ we used different if-else-clauses to be sure that we use data points of the same weekdays because of the weekly seasonality that we already considered in the naive model.\\
On the other hand we want to include the given data point exactly one week ago because we can say there is a high probability that the price is not changing that much during one week. Because we have hourly data, we have the formula
\begin{align*}
z_{3} = \text{solution}(i-168)
\end{align*}
if we want to calculate the solution at the point $i$.

\subsubsection{Include the general development in the model}
After calculating three values $z_{1}$, $z_{2}$ and $z_{3}$ we did not really consider the general development of the stock price. \\
Based on some observations, we came to the conclusion that it is a good solution to include the difference between the mean value of 2011 and the mean value of 2014 (as the general development during the last business cycle). We denote this value by z4.

\subsubsection{The resulting model}
$z_{1}$ describes the yearly seasonality while $z_{2}$ and $z_{3}$ considering the weekly seasonality and $z_{4}$ includes some kind of trend. We forecast a new data point by using these values while they are included with different weights. All in all this model is given by the formula
\begin{align*}
\text{solution}(i) = \frac{20 \cdot z1 + 70 \cdot z2 + 10 \cdot z3}{100} - \frac{z4}{2}
\end{align*}

Applied to our given data, we get the following forecast for 2015 along with errors computer:

\begin{figure}[H]
\centering \hspace{-1cm}
\includegraphics[width=12cm]{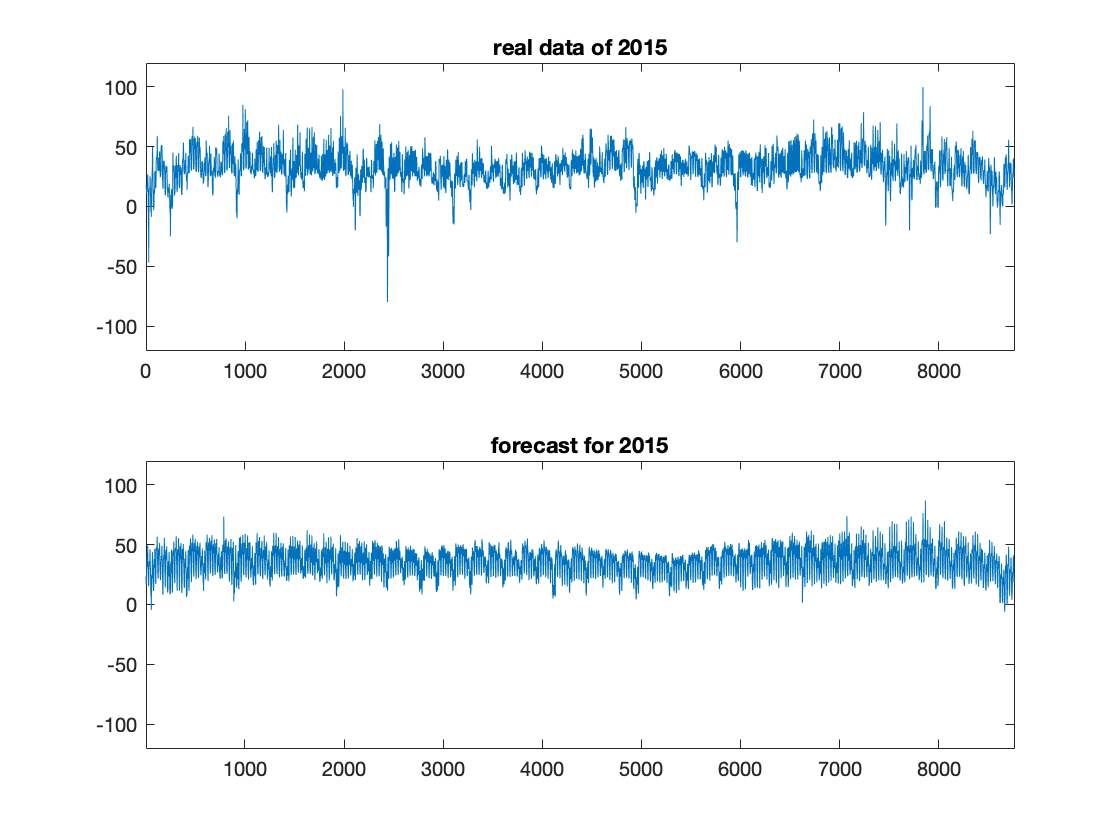}
\caption{The given data and the computed Fourier series (2015)}
\label{fig:6}
\end{figure}

\begin{figure}[H]
\centering \hspace{-1cm}
\includegraphics[width=12cm]{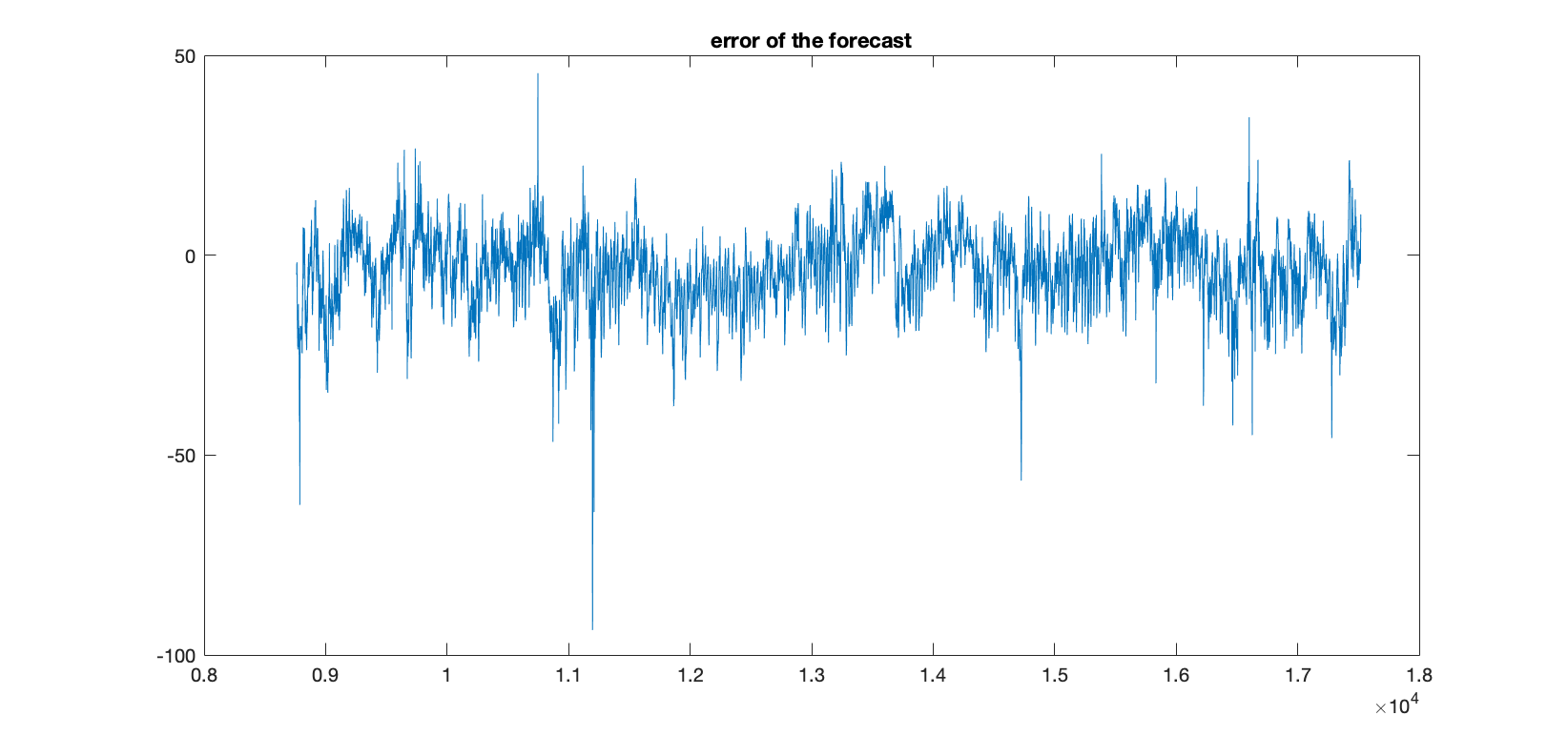}
\caption{Errors between real data and forecast}\label{fig:7}
\end{figure}

\subsection{Time Series modelling with linear predictors}\label{sec:timeseries}
One of the most conventional approaches in using time series for forecasting is the development of a linear predictor, that is, a model which represents the future values as a linear combination of the past observations. From a theoretical point of view, we assume that the time series is the realization of a certain stochastic process $X_t$. Moreover, we assume that this process can be expressed as a sum of three terms\cite{7}
\begin{math}
\end{math}

\begin{itemize}
\item \textbf{Trend} $T_t$: Trend is a deterministic function which describes general behaviour of time series.
\item \textbf{Seasonality} $S_t$: This is another deterministic function which takes into account the periodicity components of the time series.
\item \textbf{Random noise} $\epsilon_t$: A stationary\footnote{From a Mathematical point of view a stochastic process $X_t$ is stationary if $E[X_t]=\mu \ \forall t$ and the autocorrelation function between the random variables depends only on the lag $h$: $\lambda(t,s)=\lambda(t+h,s+h)$} function describing the randomness of the data.
\end{itemize}

The aim of time series modeling is to extract the random component by removing the deterministic parts. Once $\epsilon_t$ is obtained, it is possible to find a suitable model and proceed with correct forecasts.

\begin{figure}[H]
\centering
\label{subseries}
\includegraphics[scale=0.5]{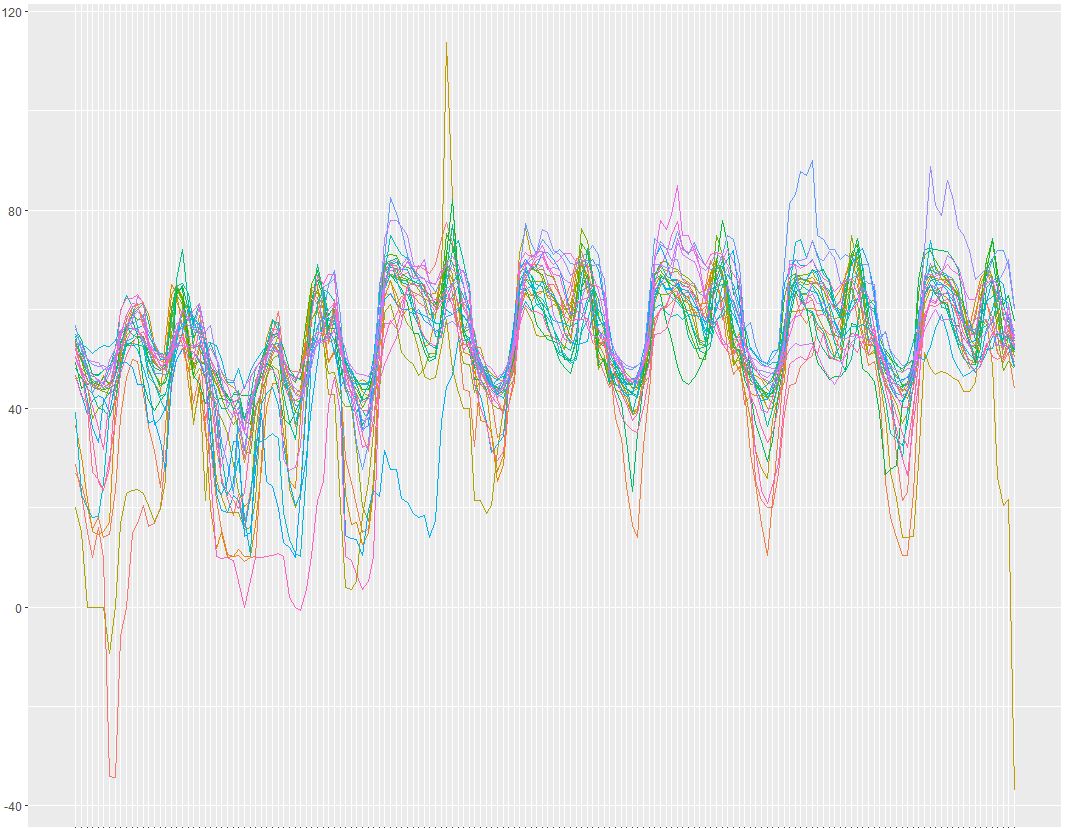}
\caption{Subseries plot of the prices over different weeks.}
\label{fig:8}
\end{figure}

Two preliminaries passages are required:
Since we are interested in long-term forecasting we take the averages of all prices during a day, so that intra-day's volatility will not influence our prevision.
First of all, is necessary to discover the periodicity of this component. There are different techniques on how to do it:
\begin{enumerate}
\item Social behaviour and economical dynamics could give us clue about a periodicity (the weekly structure deeply influences the human activity).
\item Plotting the series, or, even better, the subseries.
For example in (Fig.~\ref{fig:8}) we can see separated plots of prices from different weeks, which have the same behaviour.
\item Extracting the frequency through a Fourier Analysis. 
Indeed, we successfully applied a Fourier transform to our dataset, which highlighted the presence of a weekly seasonality.
\end{enumerate}

At this point, the following steps are applied to the dataset in order to isolate the noise:
\begin{enumerate}
\item Smoothing the data through an operation of moving averages.
\item Obtaining the seasonality component by taking averages for each day of a week.
\item Deseasonalizing the series by subtracting seasonality: $DX_t=X_t-S_t$.\item Taking differences to make series smooth and remove any residual trend:
$ \epsilon_t=DX_t-DX_{t-1}$.
\end{enumerate}

The newly obtained series $\epsilon_t$ are expected to be stationary. It is possible to verify this using some statistical test. For instance, by applying the Dickey Fuller test to the transformed data set, we confirmed that it is stationary in mean.
The stationary nature of the data is very important, because forecasting of stationary time series is supported by an important mathematical theory.

\subsubsection{Basic linear predictor}
\begin{figure}[H]
\centering
\includegraphics[scale=0.23]{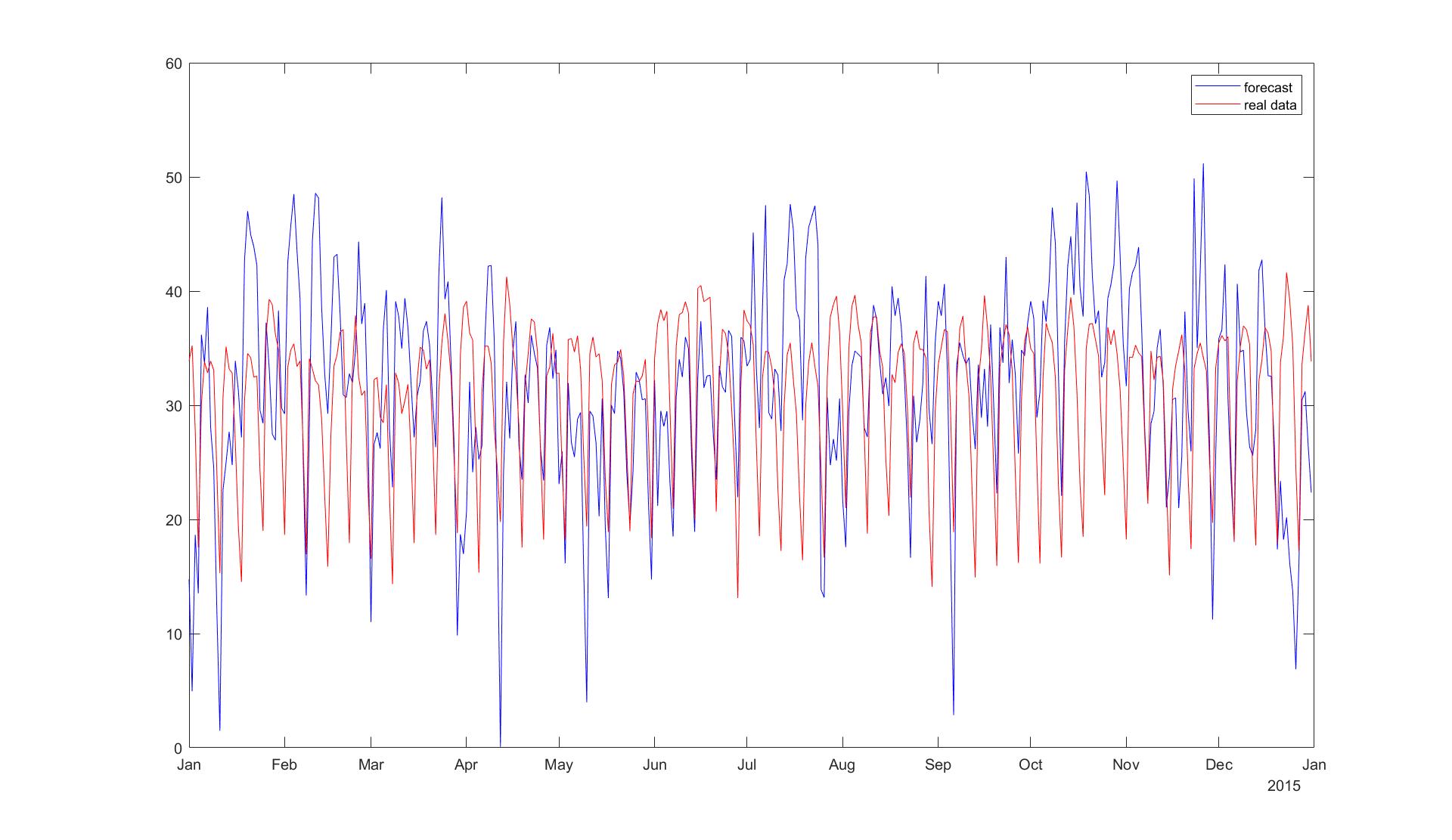}
\label{linear_predictor}
\caption{Forecast through linear predictor}\label{fig:9}
\end{figure}

The main idea in time series modelling is that the information contained in the past is used to predict the future. In other words we can express the future values as a function of the past values.In particular, we can opt for a linear model, which takes the following form: 
\begin{equation}\label{eq:basiclinear}
    \hat{X}_{n+h}=a_0 + a_1X_1 +\dots+ a_nX_n.
\end{equation}
The coefficients $a_i$ in \eqref{eq:basiclinear} have been determined by minimizing the expectation (to compute the expectation of the future value we have used the stationary property and the autocorrelation function).
The results are shown in (Fig.~\ref{fig:9}).
We can see from the picture that the model predicts the general trend of the series, but due to its simplicity it fails to fit the spikes, which are one of the most common and troublesome issues in financial forecasting. 

\subsubsection{ARMA and GARCH models}
ARMA (autoregressive moving average) models form a family of stochastic processes, whose efficiency relies on a fact that every stationary process can be represented by an ARMA process. 
Denoted by $\{Z_t\}$ a white noise process, they have the general form:
\begin{equation}\label{eq:arma}
X_t-\phi_1X_{t-1}-\dots-\phi_pX_ {t-p}=Z_t+\theta_1Z_{t-1}+\theta_qZ_{t-q}.
\end{equation}
To apply this technique, firstly, we have to choose the order of the model, i.e.\ select coefficients $p$ and $q$ in \eqref{eq:arma}, and then estimate model's parameters (this can be done automatically by MATLAB using the function \texttt{forecast}).
However, once we applied this method, the prediction turned out to be unsatisfactory: after a few number of time steps it became constant. 

\begin{figure}[H]
\centering
\label{garch}
\includegraphics[scale=0.4]{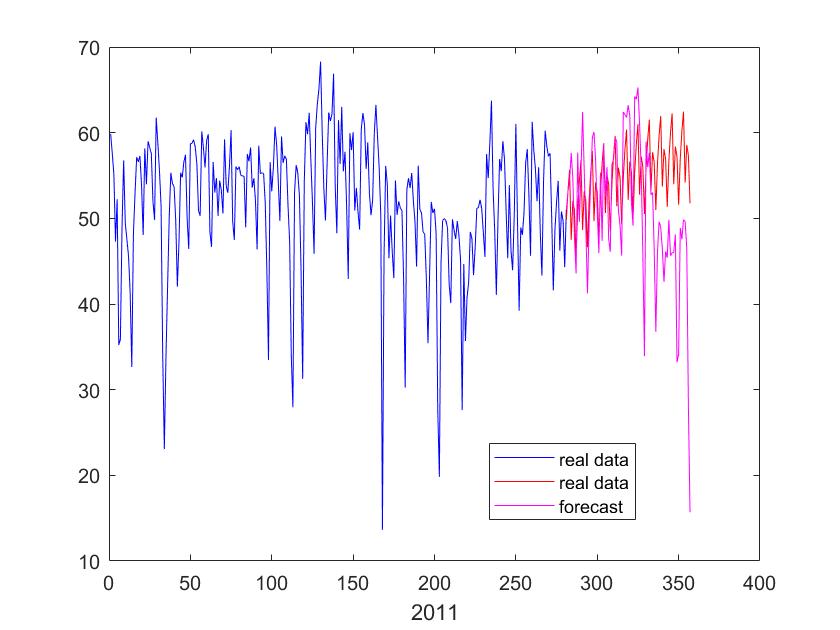}
\caption{Comparison between real data and GARCH forecastings.}
\label{fig:10}
\end{figure}

A relatively more advanced model is the GARCH (generalized autoregressive conditional heteroskedasticity). 
The peculiarity of this model is that the variance of the random term is not fixed, but is related to the previous time series's noisy terms. Thanks to this property GARCH models are employed in financial time series, because they can describe the high volatility of the assets in an efficient way. 
In (Fig.~\ref{fig:10}) we can see the result of the application of a GARCH(1,1) model. The model works quite well with the first prediction but later the forecast fails, due to the presence of a cyclic component in the data.

\subsection{Mean-reversion and jump-diffusion model (MRJD)}
\label{sec:mrjd}
The MRJD model belongs to the subgroup of reduced-form models \cite{1}. 
Reduced-form ( or quantitative, stochastic) models characterize the statistical properties of electricity prices over time. Depending on the type of market, reduced-form models can be classified as:

\begin{itemize}
    \item Spot price models, which provide a parsimonious representation of the dynamics of spot prices. 
    The two most popular sub-classes include jump-diffusion and Markov regime-switching models.As you probably guessed, we worked with exactly these prices. 
    \item Forward price models, which allow the pricing of derivatives in a straightforward manner (but only of those written on the forward price of electricity). 
\end{itemize}

One can see the MRJD model as a mixture of different processes, so let's recall the main idea behind each process. In finance, mean reversion (MR) is the assumption that a stock's price will tend to move to the average price over time. Namely, when the current market price is less than the average price, the stock is considered attractive for purchase, with the expectation that the price will rise. When the current market price is above the average price, the market price is expected to fall. In other words, deviations from the average price are expected to revert to the average. 
On the other hand, Jump-Diffusion (JD) is a stochastic process that involves jumps and diffusion. Essentially, JD is a form of mixture model, mixing a jump process and a diffusion process. A jump process (J) is a type of stochastic process that has discrete movements, called jumps, with random arrival times, rather than continuous movement, typically modelled as a simple or compound Poisson process. A diffusion process (D) is a solution to a stochastic differential equation. In our model we will use Brownian motion and Ornstein-Uhlenbeck processes, which are examples of diffusion processes \cite{2}.  

Now we are ready to get into the mathematics of our model. Electricity prices exhibit jumps in prices at periods of high demand when additional, less efficient electricity generation methods, are brought on-line to provide a sufficient supply of electricity. In addition, they have a prominent seasonal component, along with reversion to mean levels. Therefore, these characteristics should be incorporated into a model of electricity prices in the following way: 
\begin{equation}\label{eq:1}
\log(P_t) = f(t) + X_t,
\end{equation}
where $P_t$ denotes the spot price of electricity.
The logarithm of electricity price is modeled with two components: $f(t)$ and $X_t$. 
The component $f(t)$ in eqref{eq:1} is the deterministic seasonal part of the model, and $X_t$ is the stochastic part of the model. Trigonometric functions are used to model $f(t)$ in the following linear manner:
\begin{equation}\label{eq:2}
f(t) = s_1 \sin(2 \pi t) + s_2 \cos(2 \pi t) + s_3 \sin(4 \pi t) + s_4 \cos(4 \pi t) + s_5,
\end{equation} 
where $s_i, i=1,\dots,5$ are constant parameters, and $t$ 
denotes the time. The stochastic component $X_t$ is modeled as an Ornstein-Uhlenbeck process (mean-reverting) with jumps:
\begin{equation}\label{eq:3}
dX_t = (\alpha - \kappa X_t)\,dt + \sigma \,dW_t + J(\mu_J, \sigma_J) \,d\Pi(\lambda).
\end{equation}
The parameters $\alpha$ and $\kappa$ are the mean-reversion parameters. The parameter $\sigma$ is the volatility, and $W_t$ denotes a standard Brownian motion. 
The jump size is $J(\mu_J, \sigma_J)$, with a normally distributed mean $\mu_J$, and a standard deviation $\sigma_J$. 
The Poisson process $\Pi(\lambda)$ has a jump intensity of $\lambda$.

\begin{figure}[H]
\centering \hspace{-1cm}
\includegraphics[width=0.9\textwidth]{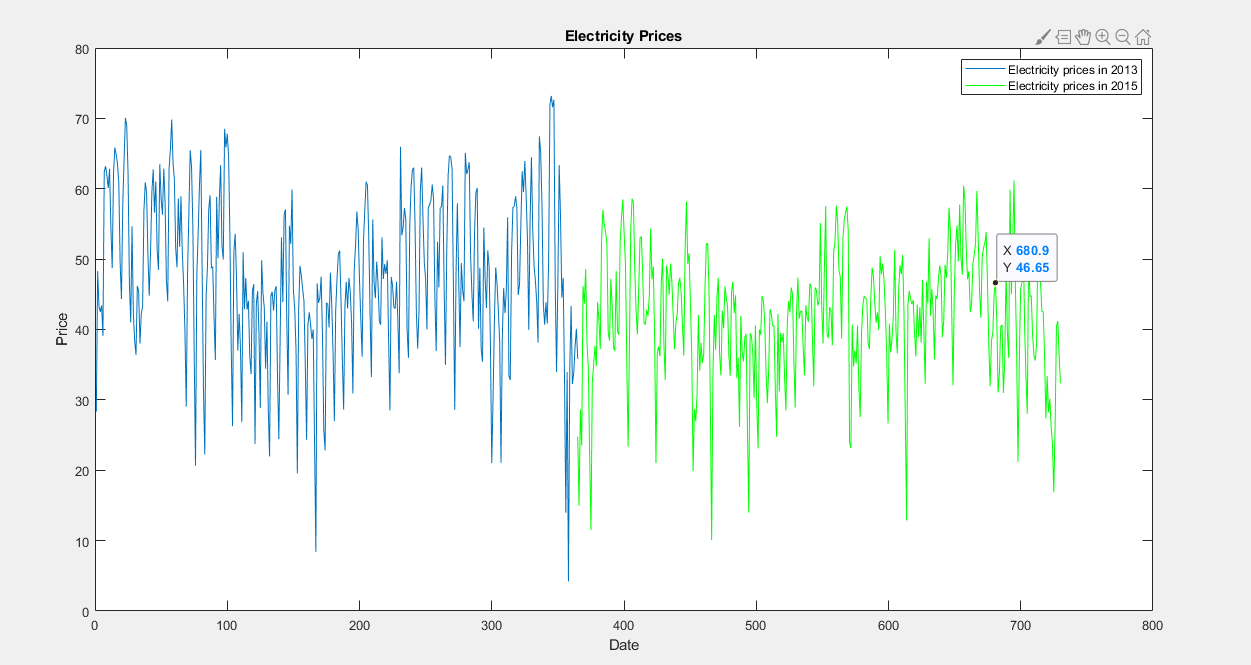} \\
\caption{Prices for 2013 and 2015 years}\label{fig:11}
\end{figure}

The electricity prices in the years 2013 and 2015 are plotted below (Fig.~\ref{fig:11}).

\subsubsection{Calibration of parameters}
The first stage was to calibrate the deterministic seasonality part using the least squares method. 
After the calibration, the seasonality is removed from the logarithm of price. 
The logarithm of price and seasonality trends are plotted below in Fig.~\ref{fig:12} 

\begin{figure}[H]
\centering \hspace{-1cm}
\includegraphics[width=0.9\textwidth]{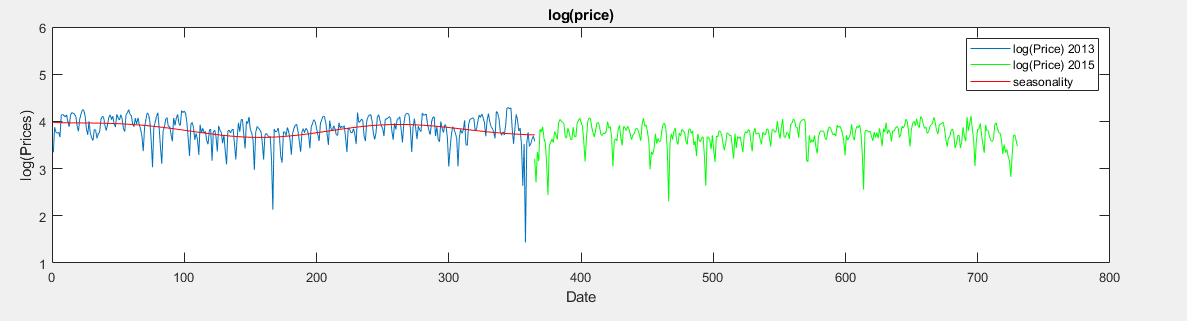}
\caption{Logarithmic prices and seasonality curve plotted}\label{fig:12}
\end{figure}

The second stage is to calibrate the stochastic part. The model for $X_t$ needs to be discretized to conduct the calibration. To discretize, assume that there is a Bernoulli process for the jump events. That is, there is at most one jump per day since this example is calibrating against daily electricity prices. The discretized equation is:
\begin{equation}\label{eq:4}
X_t=\alpha \Delta t + \phi X_{t-1} + \sigma \xi, 
\end{equation} 
with probability $(1 - \lambda \Delta t)$ and
\begin{equation}\label{eq:5}
X_t=\alpha \Delta t + \phi X_{t-1} + \sigma \xi + \mu_J + \sigma_J \xi_J,
\end{equation} 
with probability $\lambda \Delta t$, where $\xi$  and $\xi_J$ are independent standard normal random variables, 
and $\phi = 1 - \kappa \Delta t$. 
		
Finally, the parameters 
$\theta =  \{\alpha,\phi,\mu_J,\sigma^2, \sigma_J^2,\lambda\}$  can be calibrated by minimizing the negative log likelihood function:
\begin{equation}\label{eq:6}
\min_{\theta}  -\sum^T_{t=1} \log(f(X_t|X_{t-1})) \
\quad\text{subject to} \quad\phi < 1,\, \sigma^2 > 0,\, \sigma_J^2 > 0,\, 0 \leq \lambda \Delta t \leq 1
\end{equation}

\subsubsection{Forecasting} 
\begin{figure}[H]
\centering \hspace{-1cm}
\includegraphics[width=0.9\textwidth]{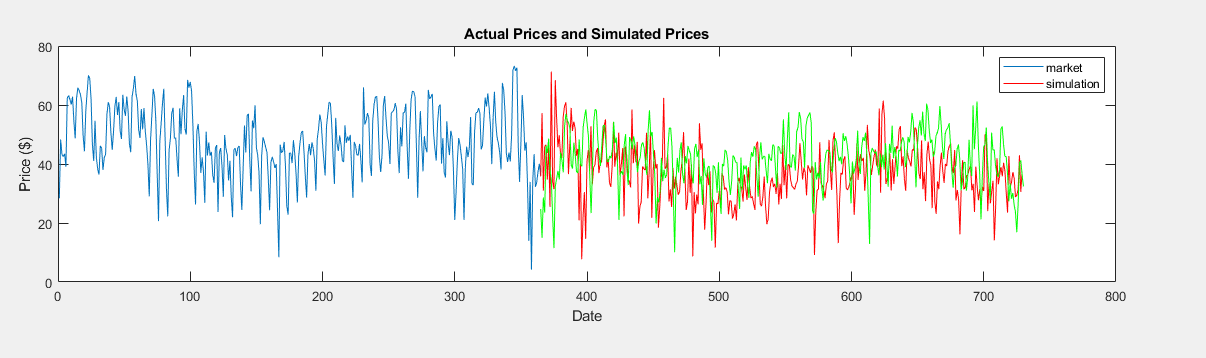} \\
\caption{The comparison between real (in green) and predicted (in red) 2015 data }\label{fig:13}
\end{figure}

The calibrated parameters and the discretized model allow us to simulate the electricity prices using the Monte Carlo method (Fig.~\ref{fig:13}).

\subsection{Artificial neural networks (ANN)}\label{sec:ann}
In the previous section, we presented the linear predictor as a prediction method. 
We now want to introduce a nonlinear predictor that also take more general input parameters into consideration. One approach is to use artificial neural networks in order to discover patterns and dependencies in the historical data (training data) and extrapolate it in order to deliver predictions.In a general formulation we search for a function $f$ such that:
\begin{equation}
\hat{{Y}}_{d,h}=f(\bm{Y}_{d-1,:}, \bm{Y}_{d-2,:},\dots,d,h,\bm{p}_{d,h},\dots).
\end{equation}
and $\hat{{Y}}_{d,h}$ must be as close as possible to the future realization of the price. 
One advantage is the possible inclusion of additional parameters (denoted by $\bm{p}_{d,h}$) that are considered to have a influence on the prediction. In this work, these parameters are the network load forecast, the day of the week, the price from the past and the month. 
In the following we give a brief introduction to artificial neural networks and then we present the application to our problem and the results.

\subsubsection{Feed forward artificial neural network}
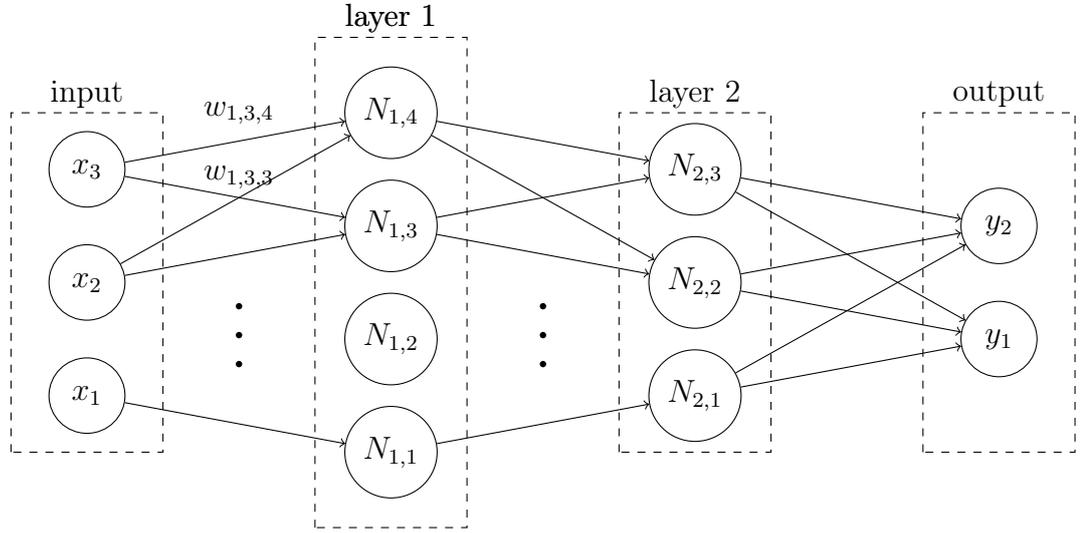
\begin{figure}[H]
	\centering
\begin{tikzpicture}
\node[circle,draw, minimum size=1cm] (N1) at  (-1,0.75) {$x_1$};
\node[circle,draw, minimum size=1cm] (N2) at  (-1,2.25)  {$x_2$};
\node[circle,draw, minimum size=1cm] (N3) at  (-1,3.75)  {$x_3$};
\node[circle,draw, minimum size=1cm] (N4) at  (3,0) {$N_{1,1}$};
\node[circle,draw, minimum size=1cm] (N5) at  (3,1.5)  {$N_{1,2}$};
\node[circle,draw, minimum size=1cm] (N6) at  (3,3)  {$N_{1,3}$};
\node[circle,draw, minimum size=1cm] (N7) at  (3,4.5)  {$N_{1,4}$};
\node[circle,draw, minimum size=1cm] (N8) at  (7,0.75) {$N_{2,1}$};
\node[circle,draw, minimum size=1cm] (N9) at  (7,2.25)  {$N_{2,2}$};
\node[circle,draw, minimum size=1cm] (N10) at  (7,3.75)  {$N_{2,3}$};
\node[circle,draw, minimum size=1cm] (N11) at  (11,1.5) {$y_1$};
\node[circle,draw, minimum size=1cm] (N12) at  (11,3)  {$y_2$};
  \draw [dashed] (-2,0) rectangle (0,4.5);
   \draw [dashed] (2,-1) rectangle (4,5.5);
    \draw [dashed] (6,0) rectangle (8,4.5);
      \draw [dashed] (10,0) rectangle (12,4.5);
\draw (1.0,4.5) node {$w_{1,3,4}$};
\draw (1.0,3.65) node {$w_{1,3,3}$};
\draw (3.0,5.75) node {layer 1};
\draw (3.0,5.75) node {layer 1};
\draw (7.0,4.75) node {layer 2 };
\draw (-1.0,4.75) node {input};
\draw (11.0,4.75) node {output};
\draw [->,above,midway] (N1) edge (N4);
\draw [->] (N3) edge (N7);
\draw [->] (N3) edge (N6);
\draw [->] (N2) edge (N6);
\draw [->] (N2) edge (N7);
\draw [->,above,midway] (N7) edge (N9);
\draw [->] (N7) edge (N10);
\draw [->] (N6) edge (N9);
\draw [->] (N6) edge (N10);
\draw [->] (N4) edge (N8);

\draw [->] (N8) edge (N11);	\draw [->] (N8) edge (N12);
\draw [->] (N9) edge (N11);	\draw [->] (N9) edge (N12);
\draw [->] (N10) edge (N11);	\draw [->] (N10) edge (N12);

\path (1,0.25) -- (1,3) node [ font=\Huge, midway, sloped] {$\dots$};
\path (5.0,0.25) -- (5.0,3) node [ font=\Huge, midway, sloped] {$\dots$};
\end{tikzpicture}
\caption{Artificial neural network with one hidden layer}
\label{fig:14}
\end{figure}

We now present the concept of the feed forward neural network. Inspired from the biological brain, a neural network mimics the propagation of the nerve impulse from neuron to neuron. The artificial neurons (also called perceptrons) are grouped in layers (see Fig.~\ref{fig:14}) and the information is propagated from the output of the neurons from one layer to the inputs of the neurons in the next layer in a fully-associative way (see Fig.~\ref{fig:14}). At the input of each neuron, the signal incoming signals are multiplied with corresponding weights and added, furthermore a bias is added. The result is then passed to an activation function.

Let $ x_{i,j}$ denote the output of the neuron $j$ from the layer $i$, and $h_i$ the activation function from the layer $i$ (neurons in the same layer have the same activation function). 
The input output characteristic has the following form:
\begin{equation}
x_{i+1,k}=h_i(w_{i,k,0}+\sum_j w_{i,k,j} \, x_{i,j}),
\end{equation}
where the $w_{i,k,j}$ is the weight of the connection between the neuron $j$ from the layer $i-1$ and neuron $k$ from the layer $i$. Notable is $w_{i,k,0}$ which is the bias. 
An input is propagated forward from the input layer $i=0$ to the output layer $\bm{x}_{N_l,:}$. 
By doing this, a general function can be approximated by a neural network with the general output function $\bm{y}=f(\bm{x},\bm{W})$.

There is an important result that states: that every continuous function defined on a bounded domain can be approximated using a neural network with one hidden layer having a non-constant, bounded and continuous activation function and one linear output layer (linear activation function). 
However learning the weights and determining the topology of the network is a tricky process. For this application, a variant of the gradient was used (Conjugate gradient backpropagation with Powell-Beale restarts).

\begin{figure}[H]
	\centering\includegraphics[width=0.9\textwidth]{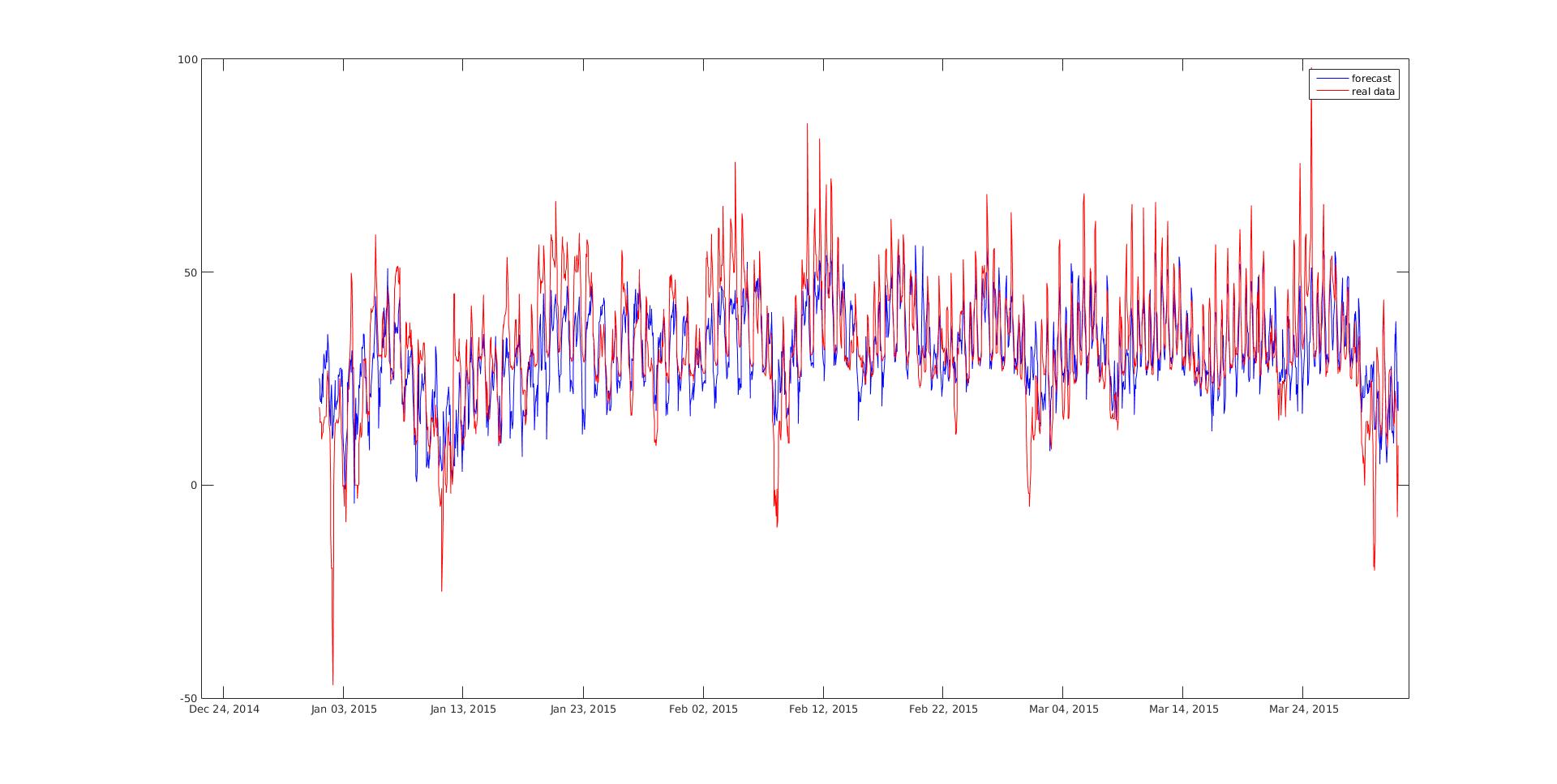}
	\caption{90 days of hourly predictions} \label{fig:16} 
\end{figure}

\begin{figure}[H]
	\centering\includegraphics[width=0.9\textwidth]{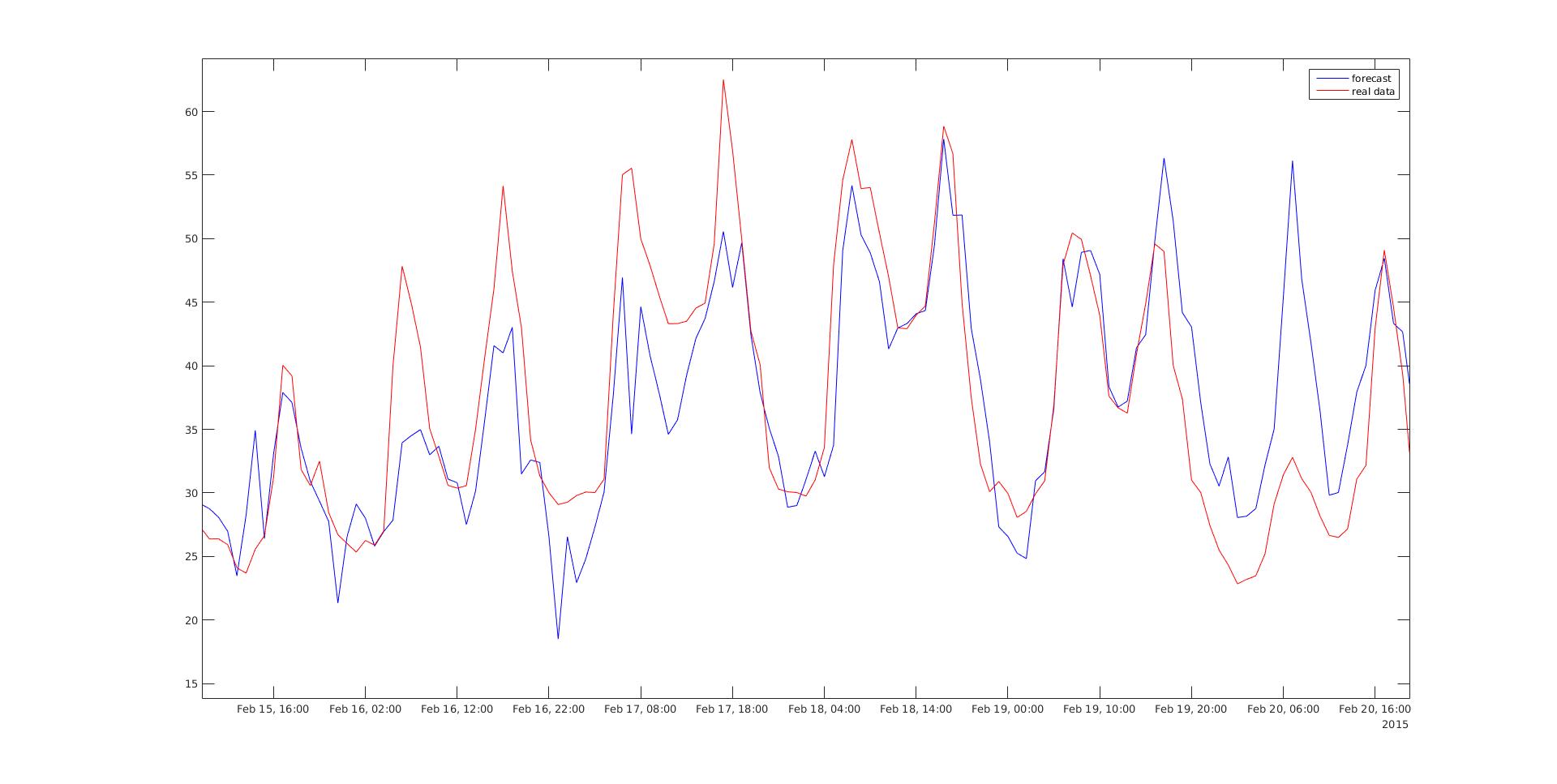}
	\caption{Zoomed prediction plot} \label{fig:17} 
\end{figure}

Determining the weights is done by optimization. The objective function that is used in this work is the mean square error:
\begin{equation}
\epsilon(\bm{W})=\frac{1}{2N_t}\sum\limits_{n=1}^{N_t}||f(\bm{x}^{(n)},\bm{W}) - \bm{y}^{(n)}||^2, \label{eq:objective}
\end{equation}
where $\{(\bm{x}^{(n)},\bm{y}^{(n)})\}_{n=1,\dots,N_t}$ is a set of $N_t$ input-output pairs that constitute the data used for training (in this application the spot price from the previous years).
\begin{figure}[H]
	\centering\includegraphics[width=0.9\textwidth]{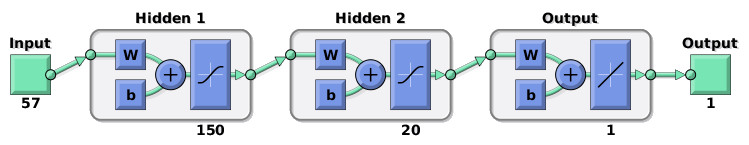}
	\caption{Structure of the used feed forward ANN} \label{fig:15} 
\end{figure}
We now apply the neural network to the energy price forecasting problem. The proposed structure is pictured in Fig.~\ref{fig:15}. The first 2 hidden layers consist of 150 respectively 20 neurons and have the sigmoid activation function. 
The output layer has only one output neuron: the network predicts the price $Y_{d,h}$ for only one hour $h$ based on values from previous days. 
The inputs are:
\begin{itemize}
	\item $\mathbf{Y}_{d-1,:} \in \mathbb{R}^7$ spot price of the previous day,
	\item $\mathbf{Y}_{d-7,:} \in \mathbb{R}^7$ spot price in the same day of the last week,
	\item $\bm{Y}_{d-14,:} \in \mathbb{R}^7$ spot price in the same day of the last 2 weeks,
	\item $\text{DoW}_d \in\{1,2,\dots,7\}$ label of the week day (Mo-Su),
	\item $t \in{0,1,\dots,23}$ hour of the prediction,
	\item $L_{d,t} \in \mathbb{R}$ load forecast,
	\item $L_{d,t}-L_{d,t-1} \in \mathbb{R}$ variation of the load.
\end{itemize}
The network is trained using the historical data from 2010-2014 and hourly predictions for 90 days are made (see Fig.~\ref{fig:16} and Fig.~\ref{fig:17}). The network is able to recognize weekly seasonality pattern as well as using the load information to predict peaks. However in some cases, the prediction has high deviations.

\subsubsection{Recurrent neural networks (RNN)}
\begin{figure}[H]
	\centering\includegraphics[width=0.9\textwidth]{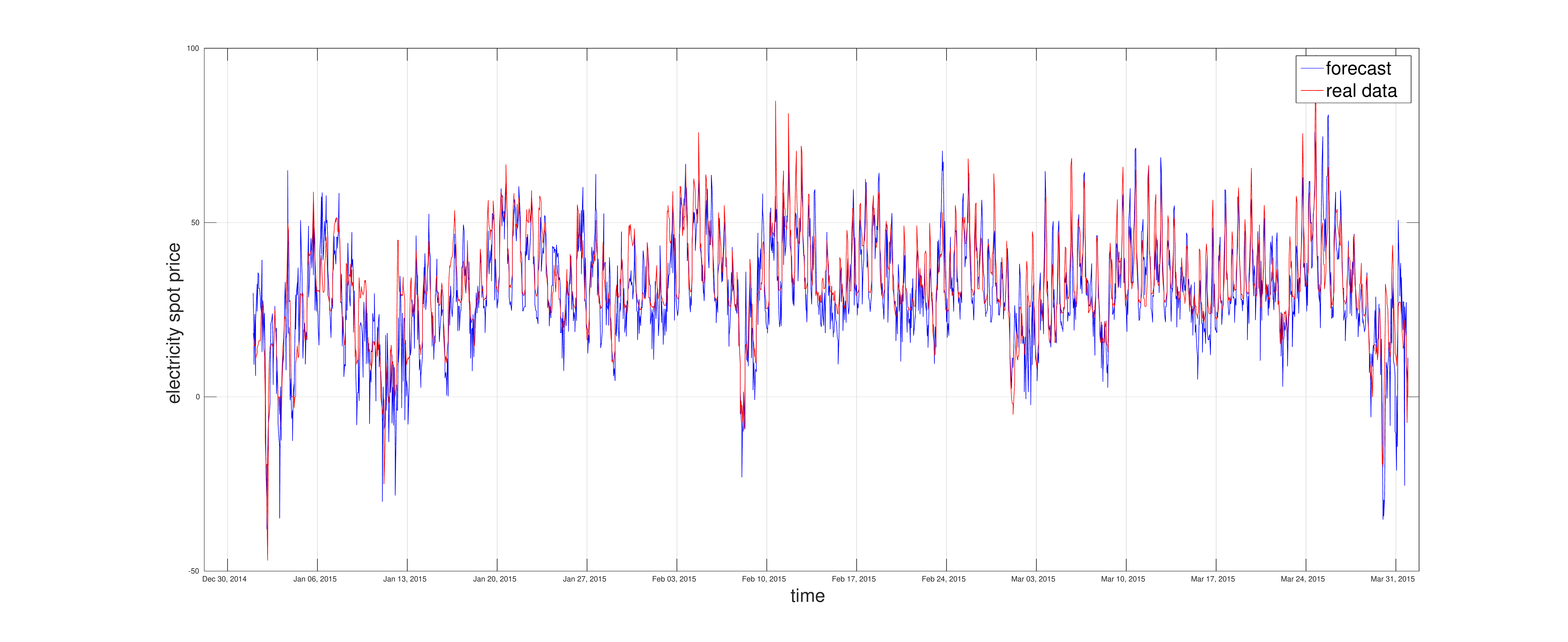}
	\caption{90 days of daily predictions using RNN} \label{fig:19} 
\end{figure}

The structure of a recurrent neural network is similar to the one presented in the previous section. 
The main difference are the feedback loops: some outputs from a a layer can be taken and stored for the next time steps and fed in as input with a time delay (see Fig.~\ref{fig:19} the feedback loop). By doing this, short term memory can be implemented. 

\begin{figure}[H]
	\centering\includegraphics[width=0.9\textwidth]{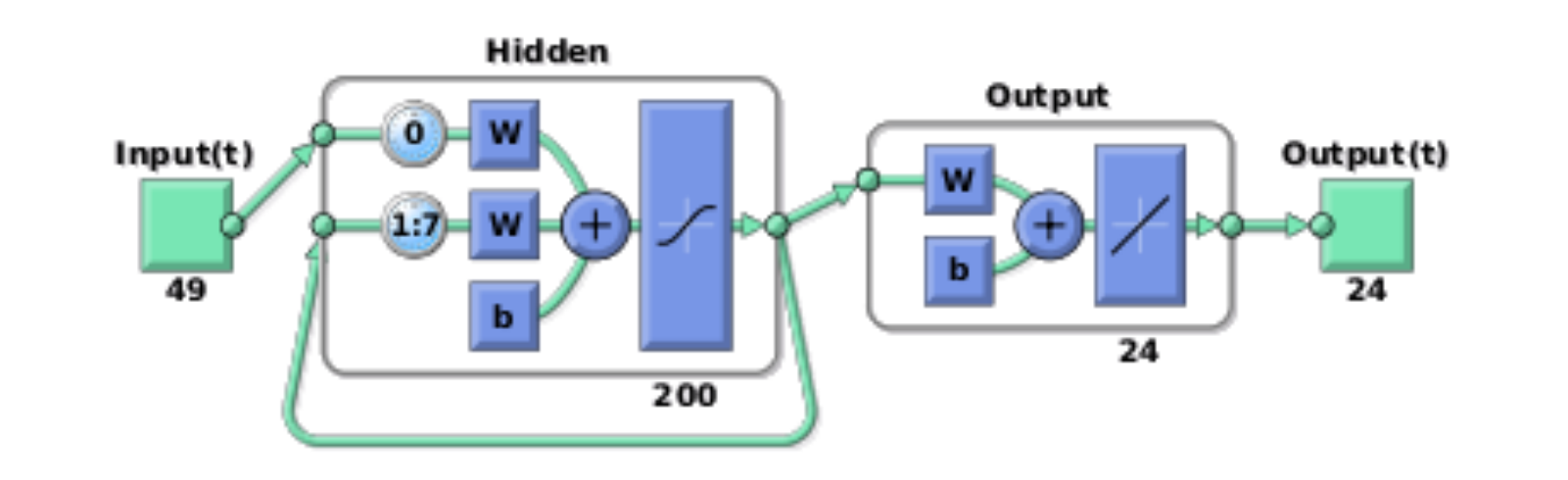}
	\caption{Structure of the used RNN} \label{fig:18} 
\end{figure}

\begin{figure}[H]
	\begin{subfigure}[b]{0.5\textwidth}
		\centering
		\includegraphics[width=\textwidth]{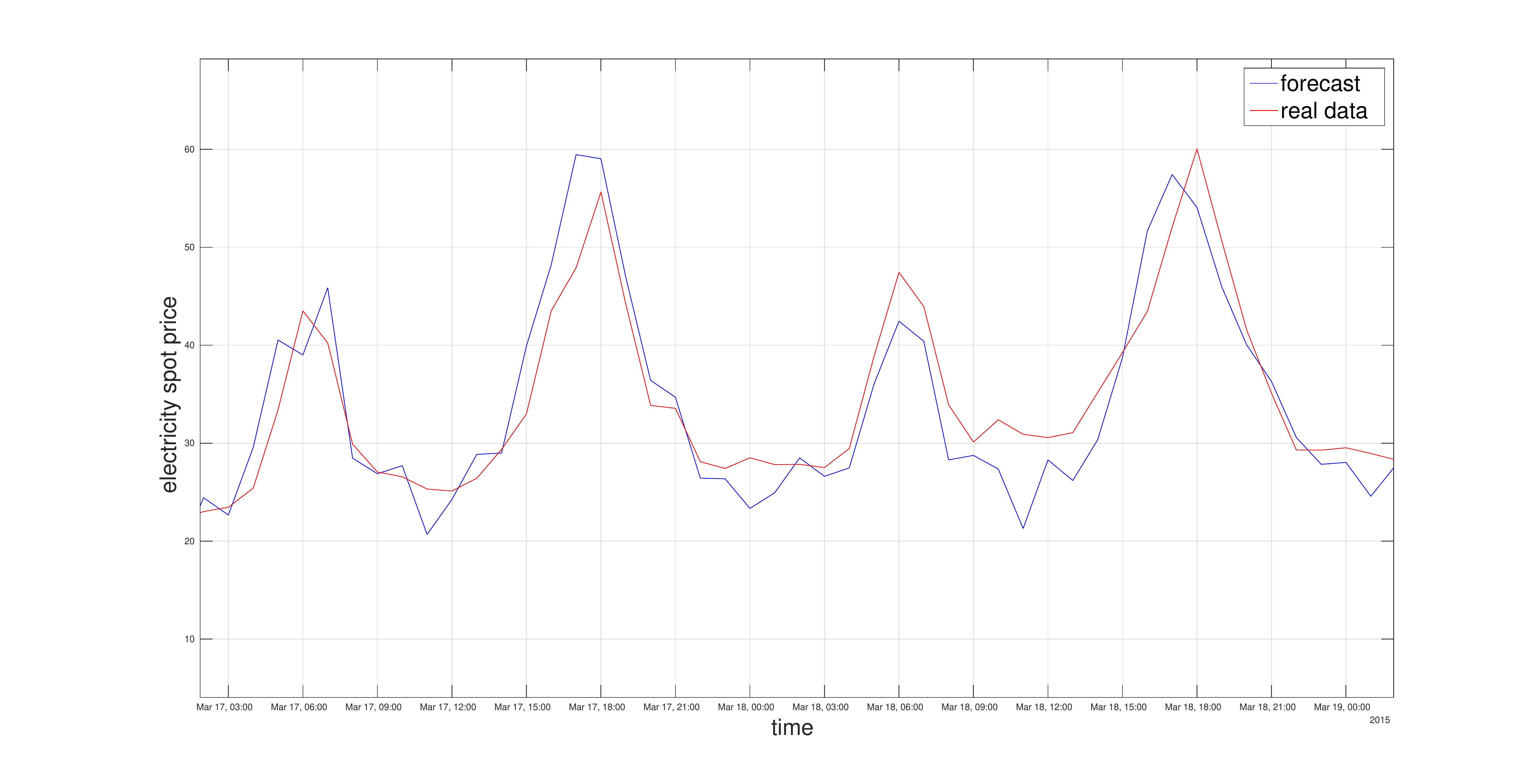}
		\caption{good predictions}
	\end{subfigure}
\begin{subfigure}[b]{0.5\textwidth}
	\centering
	\includegraphics[width=\textwidth]{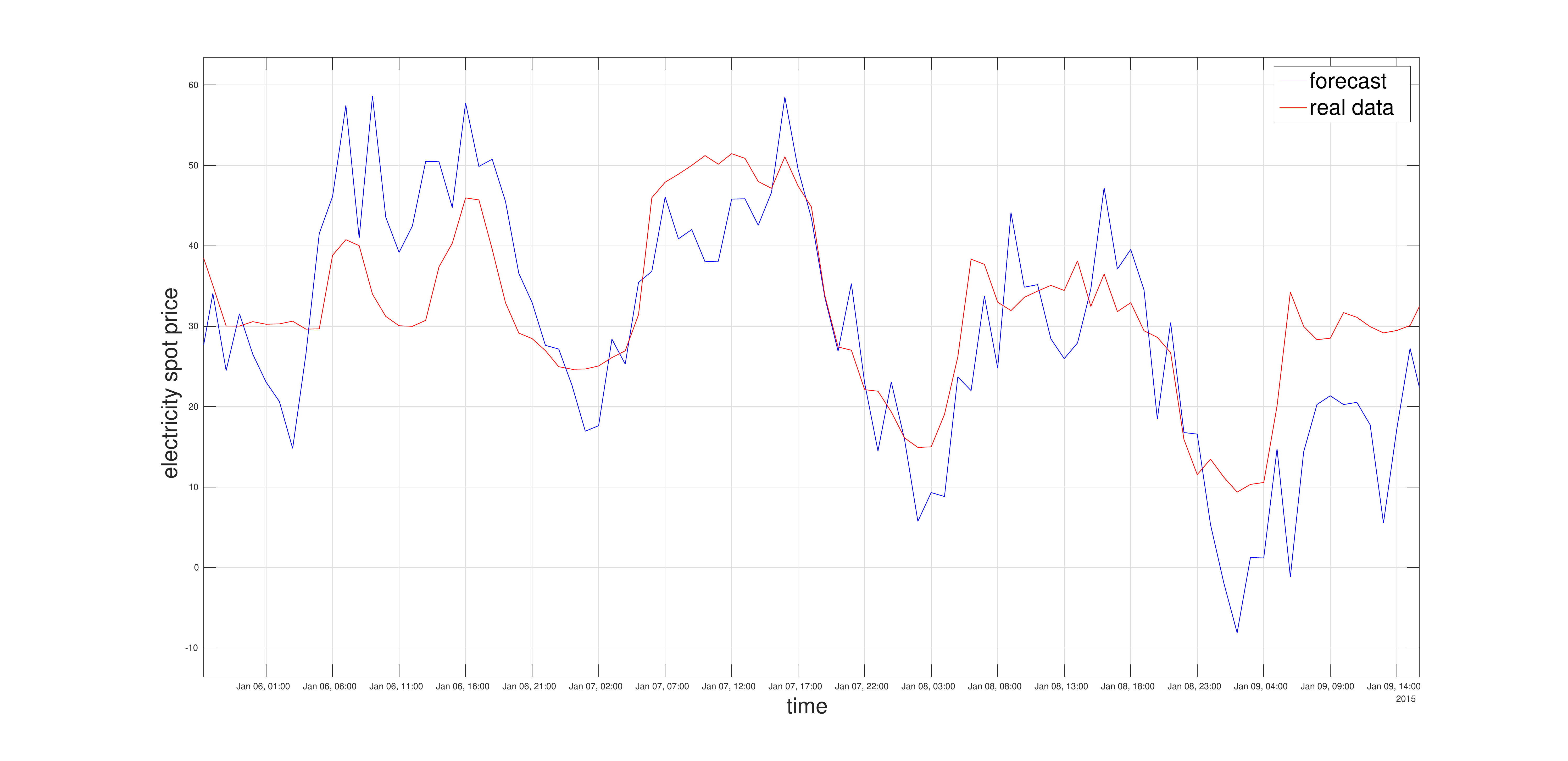}
	\caption{bad predictions}
\end{subfigure}
	\caption{examples of daily predictions using RNN}
		\label{fig:20}
\end{figure}

For the purpose of time series prediction, the following structure is proposed: a network with 2 layers, one output linear layer and one sigmoid activated layer with 200 neurons (see Fig.~\ref{fig:18}). From the output of the first layer, a feedback loop is made. The time delay is up to 7 days, The values from the past 7 days are feeded back to input with delay. 
The output represents a prediction for a whole day (predicts $\bm{Y}_{d,:}$) and the inputs are the following:
\begin{itemize}
	\item $m\in\{1,2,\dots,12\}$ - month label,
	\item $\text{DoM}\in{1,2,\dots,31}$ - day of the month,
	\item $\text{DoW}\in{1,2,\dots,7}$ - day of week,
	\item $\bm{L}_{d,:}$ - load forecast,
	\item $\bm{Y}_{d-1,:}$ - price from the previous day.
\end{itemize}
The training is done with the historical data from 2007 to 2014. A 90 day prediction from 2015 is plotted in Fig.~\ref{fig:19} and \ref{fig:20}. Compared to the normal feed forward neural network, the results suffer an improvement. 
This is due to the memory of the new neural network that allows more complicated connections to be made with a relatively low number of inputs.

\subsection{Hybrid Model}\label{sec:hybrid}
Eventually, we compared errors made in cases of MRJD, GARCH and Fourier series modelling and we decided to go with a well-known and simple hybrid method: taking averages of our 3 forecasts. 
ANA turned out to work good for hourly forecasts only, not for the yearly forecast. 
For that reason we combined only three models. The result almost replicates the price dynamics and brings us closer to the real price, by making our final forecast more accurate. 
However, this unfortunately doesn't work on every timing interval, due to several reasons which we will discuss in Section~\ref{sec:conclusions}.

\begin{figure}[H]
	\centering\includegraphics[width=0.9\textwidth]{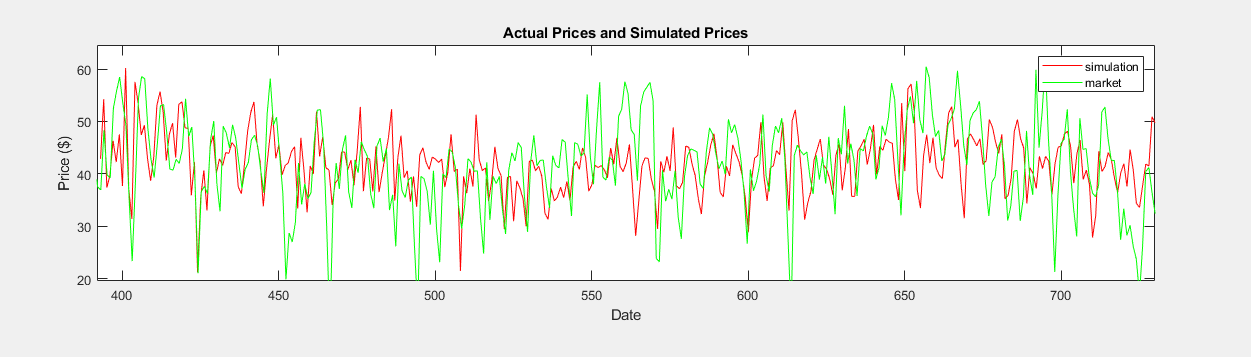}
	\caption{Forecast under the hybrid model} \label{fig:21} 
\end{figure}

\begin{figure}[H]
\centering\includegraphics[width=0.7\textwidth]{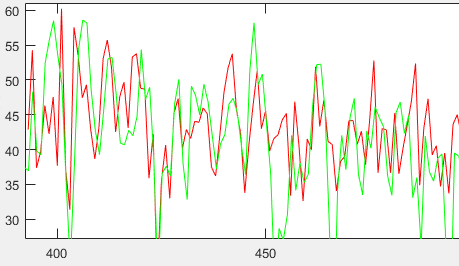}
\caption{Zoomed view} \label{fig:22} 
\end{figure}

\section{Conclusions}\label{sec:conclusions}
Starting from the Fourier Series approach, recall that to evaluate the computed forecast we plot the difference between the data points of the real data and the forecast. 
The result and errors were shown in (Fig.~\ref{fig:6}) and (Fig.~\ref{fig:7}), respectively.
As we can see the forecast includes big errors but these are caused by 
\begin{itemize}
\item the missing consideration of the noise.
\item the missing consideration of the trend (instead of calculating only a kind of trend).
\item the missing preparation of the given data with deleted outliers.
\end{itemize}
\noindent In order to fix these problems and increase the accuracy of our forecast using the Fourier series approach, it was necessary to decompose the data into a trend, seasonal component and a noise.

Linear models for time series predictions are a classical tools for forecasting and it was possible to choose among a wide range of methods. However the main difficulty is the calibration of parameters, namely, the right choice of  parameters \cite{10}.
For example for the ARMA ($p,q$) model we had to choose both the order of the model and coefficients of the linear combination. Moreover, there is not really an "absolutely correct" choice and there are only heuristic techniques to determine parameters. 
We had to proceed with many trials and errors. 
For all these reasons, it occurs that a  naive model (the basic linear predictor) performs better than a more advanced ARMA model. 
Finally, most of these methods assume that the stochastic process is stationary, an hypothesis that cannot be assured by real data. In fact  we managed to make the process stationary in mean, but it remained non-stationary in variance. 
This is linked to the high volatility of energy prices and can be partially overcame by integrating a GARCH model. Further investigations over GARCH model could be useful in order to improve the forecasting.

Regarding reduced-form method which we used, we must admit that it is not surprising that our forecast is not the most accurate ones, as on practice MRJD model usually provides simplified and inaccurate, yet realistic price dynamics. In fact, the main intention was not to provide accurate hourly price forecasts, but rather to replicate the main characteristics of daily electricity prices such as price dynamics and correlations between commodity prices \cite{9}.

As we mentioned before, the data we used was limited to past prices only, hence we didn't perform any complex data mining with a single variable. 
Nevertheless,in the end we achieved reasonable estimations with  errors being not too high. Based on the results of teams from GEFCom 2014 we concluded, however, that data mining must always be present in the ESPF process, being one of three consistent stages of forecasting. Teams were given the huge amount of data and were asked, first of all, to come up with appropriate methods for data mining. As a result, the number of variables they ended up with varied between twenty and seventy. 

Overall, despite the fact that "previous prices" based models are sometimes used for ESPF, in the common day-ahead forecasting scenario authors select some combinations of early-mentioned drivers, relying on heuristics and experience of a forecaster. Such an approach helps to achieve better results with smaller errors. 





% ...

\end{document}